\documentclass{article}
\usepackage[margin=2cm]{geometry}

\usepackage{float}

\usepackage{changepage}

\usepackage[utf8]{inputenc}

\usepackage{textcomp,marvosym}

\usepackage{fixltx2e}

\usepackage{amsmath,amssymb,amsthm}

\usepackage{cite}

\usepackage{nameref,hyperref}

\usepackage{microtype}
\DisableLigatures[f]{encoding = *, family = * }

\usepackage{rotating}

\usepackage[aboveskip=1pt,labelfont=bf,labelsep=period,justification=raggedright,singlelinecheck=off]{caption}


\DeclareMathOperator{\Cov}{Cov}
\DeclareMathOperator{\Var}{Var}

\DeclareMathOperator{\Proba}{\mathbb{P}}
\newcommand{\Covb}[2]{\ensuremath{\Cov\!\left[#1,#2\right]}}

\newcommand{\Pb}[1]{\ensuremath{\Proba\!\left[#1\right]}}
\newcommand{\Varb}[1]{\ensuremath{\Var\!\left[#1\right]}}
\newcommand{\norm}[1]{\| #1 \|}
\newcommand{\indep}{\rotatebox[origin=c]{90}{$\models$}}

\usepackage{ragged2e}

\usepackage{ragged2e}

\title{Second-order Control of Complex Systems with Correlated Synthetic Data}
\date{}

\author{Juste Raimbault\textsuperscript{1,2,3,*}\medskip\\
\textsuperscript{1}CASA, UCL, London, UK\\
\textsuperscript{2}UPS CNRS 3611 ISC-PIF, Paris, France\\
\textsuperscript{3}UMR CNRS 8504 G{\'e}ographie-cit{\'e}s, Paris, France\medskip\\
Email : juste.raimbault@polytechnique.edu}

\begin{document}

\maketitle

\begin{abstract}
The generation of synthetic data is an essential tool to study complex systems, allowing for example to test models of these in precisely controlled settings, or to parametrize simulation models when data is missing. This paper focuses on the generation of synthetic data with an emphasis on correlation structure. We introduce a new methodology to generate such correlated synthetic data. It is implemented in the field of socio-spatial systems, more precisely by coupling an urban growth model with a transportation network generation model. We also show the genericity of the method with an application on financial time-series. The simulation results show that the generation of correlated synthetic data for such systems is indeed feasible within a broad range of correlations, and suggest applications of such synthetic datasets.
\end{abstract}

\justify

\section*{Introduction}

Developing methods to study complex systems, such as simulation models or data-mining techniques, often requires testbeds and benchmarks to ensure expected properties. The use of synthetic data, in the sense of statistical populations generated randomly under constraints of proximity of patterns to a studied system, is a widely used methodology tackling this issue. This approach is used in several disciplines related to complex systems such as therapeutic evaluation~\cite{abadie2010synthetic}, territorial science~\cite{moeckel2003creating,pritchard2009advances}, machine learning~\cite{bolon2013review} or bio-informatics~\cite{van2006syntren}.

Generation of synthetic datasets can consist in data disaggregation by producing a microscopic population with fixed macroscopic properties \cite{beckman1996creating}. The creation of synthetic populations for microsimulation models is a typical example where empirical statistical distributions are reproduced \cite{muller2010population}. In data extensive contexts, several methods have been developed and improved for a better reproduction of margin distributions \cite{barthelemy2013synthetic}.

Synthetic datasets can also be generated at the same scale than the targeted real dataset, with a broad range of realism levels and corresponding constraints on the generated data \cite{hoag2008synthetic}. For example, \cite{eno2008generating} show that some datamining techniques such as decision trees can be inverted to produce datasets capturing complex non-linear patterns.

The constraints of proximity to reality of synthetic dataset will depend on expected applications. They range for example from a strong statistical fit on given indicators, to weaker assumptions of similarity on aggregated patterns. In the case of systems where emergence plays a central role, a microscopic property does not directly imply given macroscopic patterns, and synthetic datasets may have to capture some of these. This approach therein becomes part of the complex systems simulation toolbox. Indeed, with the rise of new computational paradigms~\cite{arthur2015complexity}, data (simulated, measured or hybrid) shape our understanding of complex systems. Methodological tools for data-mining, modeling and simulation, including the generation of synthetic data, are therefore crucial to be developed.

\subsection*{Synthetic data and dependancy structures}

Reproducing data patterns at the first order, in the sense of distribution moments, is broadly used and understood. A targeted average will be easily reproduced. Similarly, marginals are fitted when generating synthetic population. However, higher orders of data structure are more difficult to include in synthetic data generation methods. At the second order, this corresponds to a control of the covariance structure between generated variables.

Some specific examples where interdependency structure is controlled can be found. \cite{ye2011investigation} investigates the sensitivity of discrete choices models to the distributions of inputs and to their dependance structure. \cite{birkin1988synthesis} develop a generic framework to generate synthetic micro-data from heterogenous aggregated data sources, which in particular can include second-order effects in the models considered. \cite{li2014differentially} propose to reconstruct multi-dimensional synthetic data using copulas, which capture the dependancy structure between marginal distributions. It is also possible to interpret complex networks generative models~\cite{newman2003structure} as the production of an interdependence structure for a system, contained within link topology. Most methods yielding a high level of accuracy on synthetic covariance structure depend on sampling or data reconstruction methods, and need therefore large datasets.

\subsection*{Synthetic data and socio-spatial systems}

Synthetic data with a spatial dimension, in the sense of spatial coordinates of generated data points, or more complicated spatial structures, require proper methods and paradigms. Such approaches have been proposed in disciplines such as geostatistics or Earth sciences. \cite{robin1993cross} describe a method to generate cross-correlated random spatial fields using Fourier transforms. \cite{osborn2017multilevel} introduce a multilevel sampling technique to produce correlated random fields. Concrete applications of such spatial synthetic data include atmospheric circulation models \cite{gourdji2010regional}, rainfall-runoff simulations \cite{robin1993cross}, or engineering \cite{zhao2018simulation}.

In the case of socio-spatial systems, these kind of methods is less developed. Simulation approaches to spatialized social systems are already well studied by disciplines such as geosimulation \cite{benenson2004geosimulation}, urban analytics \cite{batty2013new} or theoretical and quantitative geography \cite{pumain2018evolutionary}. The use of synthetic data in these contexts is however systematically reduced to the generation of synthetic populations within agent-based models or microsimulation models, applied for example to mobility \cite{banos2005simulating}, land-use transport interaction models~\cite{pritchard2009advances}, or demography microsimulation models \cite{birkin1988synthesis}. Some techniques in spatial statistics, such as Geographically Weighted Regression \cite{brunsdon1998geographically}, can also be understood as extrapolating a spatial field and thus constructing spatial synthetic data.

While several examples of stylized models initialized on synthetic configurations can be found in the literature, such as the first Simpop model~\cite{sanders1997simpop} to simulate the dynamics of settlements at a macroscopic scale, or the SimpopNet model \cite{schmitt2014modelisation} for the co-evolution of cities and transportation networks, these are run on a single stylized synthetic configuration. There is to the best of our knowledge very few examples of works coupling a synthetic data generator with a model at an other scale than the microscopic scale of the population. 

Recently, a systematic control of the effects of the initial spatial configuration on the behavior of simulation models was proposed by~\cite{raimbault2018space}. The aim is to be able to distinguish proper effects due to intrinsic model dynamics from particular effects due to the geographical structure of the case study. \cite{arentze2012modeling} introduce a method to generate realistic social networks associated to a synthetic population in the geographical space. Such results are essential for the validation of conclusions obtained with modeling and simulation practices in quantitative geography. Being able to generate correlated synthetic configurations of territorial systems is thus an important development remaining to be investigated. In such systems, spatio-temporal correlation structures are a proxy to capture complex dynamics, and controlling them in synthetic data would allow better understanding of models of such systems.

\subsection*{Proposed approach}

This literature review on different aspects of synthetic data generation unveils at least two gaps: (i) a lack of attention on controlling covariance structures when generating synthetic data; and (ii) an absence of such methods applied to the study of socio-spatial systems at aggregated scales. As spatio-temporal dependencies structures are essential in driving the dynamics of such systems \cite{pigozzi1980interurban,chen2009urban}, the combination of these two aspects appears as an unexplored research problem.

We propose in this paper to study the generation of correlated synthetic data, and more particularly in the case of socio-spatial systems. We introduce here a generic methodology taking into account the dependance structure for the generation of synthetic datasets, more precisely by controlling the average of correlation matrices. It is suited to be applied on cases where microscopic data is not available and system similarity is expected on aggregated indicators.

We investigate thus the question of how to generate correlated synthetic data at aggregated levels, where constraints on macroscopic indicators are fulfilled and correlation structure is controlled. We focus on this problem in the particular case of socio-spatial systems, but keep in mind the genericity of the approach.

Our contribution is twofold: (i) we implement a generation of spatial synthetic data for socio-spatial systems, which to the best of our knowledge has never been done in that context; (ii) the method introduced is generic, and we illustrate it with an application to financial time-series.

The rest of the paper is organized as follows. The generic method to generate correlated synthetic data is first formally described. We then apply it to a generative model of territorial configurations, composed by the sequential coupling of a reaction-diffusion model for population density with a road network generation model, and study the produced correlation patterns. We also illustrate in a following section illustrate the genericity of our method by applying it to financial time-series, which are an other example of highly complex signals for which correlations are crucial.

\section*{Method Formalization}

The domain-specific methods described above are too broad to be summarized within a same formalism. We therefore introduce here a generic and model-agnostic framework, focused on the control of correlations structures in synthetic data.

Let $\vec{X}_I$ a multidimensional stochastic process (which can be indexed e.g. with time in the case of time-series, but also with space, or any other indexation). We assume to have a real dataset $\mathbf{X}=(X_{i,j})$, which is interpreted as a set of realizations of the stochastic process. We propose to generate a statistical population $\mathbf{\tilde{X}}=\tilde{X}_{i,j}$ such that
\begin{enumerate}
\item A given criteria of proximity to data is verified, i.e. given a precision $\varepsilon$ and some aggregated indicator $\vec{f}$, we have 
\begin{equation}
\label{eq:data-proximity}
\norm{\vec{f}(\mathbf{X})- \vec{f}(\mathbf{\tilde{X}})} < \varepsilon
\end{equation}
\item The level of correlation is controlled, i.e. given a matrix $\mathbf{R}$ representing the correlation structure (any symmetric matrix with coefficients in $[-1,1]$ and a unity diagonal), we have the estimated covariance matrix given by
\begin{equation}
\hat{\Cov{}{}}\left[\mathbf{\tilde{X}}\right] = \mathbf{\Sigma}^{T} \cdot \mathbf{R} \cdot \mathbf{\Sigma}
\end{equation}
where the standard deviation diagonal matrix $\mathbf{\Sigma}$ is estimated on the synthetic population.
\end{enumerate}

The second requirement will generally be conditional to parameter values determining generation procedure, either generation models being simple or complex ($\mathbf{R}$ itself is a parameter). Formally,  we can also understand synthetic processes as parametric families $\tilde{X}_i[\vec{\alpha}]$.

We propose to apply the methodology on very different examples, both typical of complex systems: territorial systems and financial high-frequency time-series. We illustrate the flexibility of the method, and claim to help building interdisciplinary bridges by methodology transposition and reasoning analogy. In the first case, morphological calibration of a population density distribution model allows to respect real data proximity. Correlations of urban form with transportation network measures are empirically obtained by exploration of coupling with a network morphogenesis model. The control is in this case indirect and the feasible space of correlations is empirically determined. In the second case, proximity to data is the equality of signals at a fundamental frequency, to which higher frequency synthetic components with controlled correlations are superposed.

\section*{Correlated population density and road network}

We now apply the method to territorial systems of human settlements, in the particular case here of population distribution in correlation with road network. In this application, simulation appears as a crucial step to implement the method.

\subsection*{Territorial configuration model}

We propose in our case to generate territorial systems summarized in a simplified way as a spatial population density $d(\vec{x})$ and a transportation network $n(\vec{x})$. Correlations we aim to control are correlations between urban morphological measures and network measures. The question of interactions between territories and networks is already well-studied~\cite{offner1996reseaux} but remains highly complex and difficult to quantify~\cite{offner1993effets}. A dynamical modeling of implied processes should shed light on these interactions \cite{bretagnolle:tel-00459720}, and \cite{raimbault2018caracterisation} has investigated the concept of co-evolution within such models. We develop here in that context a simple coupling (i.e. without any feedback loop) between a population density distribution model and a network morphogenesis model.

\subsubsection*{Density model}

We use a model $D$ similar to aggregation-diffusion models~\cite{batty2006hierarchy} to generate a discrete spatial distribution of population density. A generalization of the basic model is proposed in~\cite{raimbault2018calibration}, providing a calibration on morphological objectives (entropy, hierarchy, spatial auto-correlation, mean distance) against real values computed on the set of 50 km sized grid extracted from European density grid~\cite{eurostat}. We recall here rapidly the processes included in the model. An square grid of width $W$, initially empty, is represented by population $(P_i(t))_{1\leq i\leq W^2}$. At each time step, until the total population reaches a fixed parameter $P_m$,
\begin{itemize}
\item total population is increased of a fixed number $N_G$ (growth rate), following a preferential attachment such that 
\begin{equation} 
\Pb{P_i(t+1)=P_i(t)+1|P(t+1)=P(t)+1}=\frac{(P_i(t)/P(t))^{\alpha}}{\sum(P_i(t)/P(t))^{\alpha}}
\end{equation}
\item a fraction $\beta$ of population is diffused to four closest neighbors is operated $n_d$ times
\end{itemize}

The two opposite processes of urban concentration and urban sprawl are captured by the model, what allows reproducing with a good precision a large number of existing morphologies regarding macroscopic urban form indicators.

\subsubsection*{Network model}

On top of the population density model, we generate a planar transportation network with a model $N$ at a similar scale. Several processes can be taken into account to simulate network growth \cite{raimbault2018multi}. Other model types could be used as well, such as biological self-generated networks~\cite{tero2010rules}, local network growth based on geometrical constraints optimization~\cite{courtat2011mathematics}, or a more complex model based on multi-dimensional network percolation \cite{raimbault2019multi} which would allow the creation of loops for example. \cite{raimbault2018multi} generates networks in the frame of a modular architecture, in which the choice of the network generation heuristic can be adapted to a specific need (as e.g. proximity to real data, constraints on output indicators, variety of generated forms, etc.).

We choose here an heuristic based on spatial interaction potential breakdown, which corresponds in practice to a network answering to the strongest demand patterns. The algorithm assumes realistic thematic assumptions: a connected initial network and the creation of links based on spatial interactions.

The heuristic network generation procedure is the following:
\begin{enumerate}
\item A fixed number $N_c$ of centers that will be first nodes of the network are distributed given density distribution. Their spatial distribution follows a similar law than the aggregation process, i.e. the probability to be distributed in a given patch is $\frac{(P_i/P)^{\alpha}}{\sum (P_i/P)^{\alpha}}$. Population is then attributed according to Voronoi areas of centers, such that a center cumulates population of patches within its triangulation extent.
\item Centers are connected deterministically through a percolation between closest clusters: as long as the network is not fully connected, the two closest connected components, in the sense of the minimal euclidian distance between their vertices, are connected with the link realizing this distance. It yields a tree-shaped network at this stage.
\item The network is modified by adding links following a spatial interaction potential breaking. More precisely, a generalized gravity potential between two centers $i$ and $j$ is defined by
\begin{equation}
V_{ij}(d) = \left[ (1 - k_h) + k_h \cdot \left( \frac{P_i P_j}{P^2} \right)^{\gamma} \right]\cdot \exp{\left( -\frac{d}{r_g (1 + d/d_0)} \right)}
\end{equation}
where $d$ can be euclidian distance $d_{ij}=d(i,j)$ or network distance $d_N(i,j)$, $k_h \in [0,1]$ is a weight to determine the role of populations, $\gamma$ gives the shape of the hierarchy across population values, $r_g$ is a characteristic interaction distance and $d_0$ is a distance shape parameter.
\item A fixed number $K\cdot N_L$ of potential new links is taken among the couples having greatest euclidian distance potential ($K=5$ is fixed).
\item Among potential links, $N_L$ are effectively realized, that are the one with smallest rate $\tilde{V}_{ij} = V_{ij}(d_N)/V_{ij}(d_{ij})$. At this stage only the gap between euclidian and network distance is taken into account: $\tilde{V}_{ij}$ does indeed not depend on populations and is increasing with $d_N$ at constant $d_{ij}$.
\item Planarity of the network is imposed by creating nodes at possible intersections formed by new links.
\end{enumerate}

The nature and range of correlations produced by this model coupling, as a function of model parameters, are to be determined by simulation experiments.

\subsubsection*{Parameter space}

The parameter space for the coupled model is constituted first by density generation parameters $\vec{\alpha}_D = (P_m/N_G , \alpha,\beta , n_d)$. We study for the sake of simplicity the rate between population and growth rate $P_m/N_G$ instead of both varying, i.e. the number of steps needed to generate the distribution. These are completed by network generation parameters $\vec{\alpha}_N=(N_C,k_h,\gamma , r_g , d_0)$. We write $\vec{\alpha} = (\vec{\alpha}_D,\vec{\alpha}_N)$.

\subsubsection*{Indicators}

Urban form and network structure are quantified by numerical indicators in order to modulate correlations between these. Morphology is defined as a vector $\vec{M}=(r,d,\varepsilon,a)$ giving spatial auto-correlation (Moran index), mean distance, entropy and hierarchy (see~\cite{le2015forme} for a precise definition of these indicators). Network measures $\vec{G} = (c,l,s,\delta)$ are with network denoted $(V,E)$
\begin{itemize}
\item Average centrality $c$ defined as average \emph{betweeness-centrality} (normalized in $[0,1]$) on all links.
\item Average path length $l$ given by $\frac{1}{d_m}\frac{2}{|V|\cdot (|V|-1)}\sum_{i<j}d_N(i,j)$ with $d_m$ normalization distance taken here as world diagonal $d_m=\sqrt{2}N$.
\item Average network speed~\cite{banos2012towards} which corresponds to network performance compared to direct travel, defined as $s = \frac{2}{|V|\cdot (|V|-1)}\sum_{i<j}{\frac{d_{ij}}{d_N(i,j)}}$.
\item Network diameter $\delta = \max_{ij}d_N(i,j)$.
\end{itemize}

We study the cross-correlation matrix $\Covb{\vec{M}}{\vec{G}}$ between morphology and network. We estimate it on a set of $n$ realizations at fixed parameter values $(\vec{M}\left[D(\vec{\alpha})\right],\vec{G}\left[N(\vec{\alpha})\right])_{1\leq i\leq n}$ with the standard unbiased estimator, given by Eq.~\ref{eq:correstimate} below.

\begin{equation}
\label{eq:correstimate}
\hat{\rho}[X1,X2] = \frac{\hat{C}[X1,X2]}{\sqrt{\hat{\Var{}}[X1] \cdot \hat{\Var{}}[X2]}}
\end{equation}

The covariance is estimated by Eq.~\ref{eq:covariance}, where variables are indexed by $t$ over $T$ realizations.

\begin{equation}
\label{eq:covariance}
\hat{C}[X1,X2] = \frac{1}{(T-1)}\sum_{t} X_1(t)X_2(t) - \frac{1}{T\cdot (T-1)} \sum_t X_1(t) \sum_t X_2(t)
\end{equation}

The variance is estimated by Eq.~\ref{eq:variance}.

\begin{equation}
\label{eq:variance}
\hat{\Var{}}[X] = \frac{1}{T}\sum_t{X^2(t)}-\left(\frac{1}{T}\sum_tX(t)\right)^2
\end{equation}

\subsubsection*{Null model}

In order to provide a minimal benchmark of our correlated data generation method, we also introduce a null model to control if the produced correlation are not intrinsic to the specification of indicators for example. The procedure to generate null configuration is the following: (i) generate a random population density, by randomly selecting a proportion $r_o^{(0)}$ of occupied cell and attributing them a random density between 0 and 1; (ii) add a fixed number of network nodes $N_N^{(0)}$, either randomly in space, or following the population density with a probability of each cell proportional to its density; (iii) add a fixed number of links $N_L^{(0)}$ between random pairs of nodes; (iv) planarize the resulting network by adding nodes at link intersections. In this model, population density and network are either totally independent, or linked through network node density only. We thus expect the corresponding correlation to behave as a baseline of how correlations between indicators behave when no particular care is given to including interaction processes.

\subsection*{Generating correlated synthetic data}

The coupling of generative models is done both in a formal and operational way. We indeed loosely couple independent implementations. The OpenMOLE software~\cite{reuillon2013openmole} for model exploration offers a proper framework for this. Its modular workflow language allows to compose model tasks and integrate these into diverse numerical experiments. For the population density generation, we use the \texttt{scala} implementation provided by \cite{raimbault2018calibration}. The network generation model is implemented in NetLogo~\cite{wilensky1999netlogo}, which offers a good compromise between performance and interactive model validation and exploration. The two models are coupled with a specific OpenMOLE script. Source code is available at \url{https://github.com/JusteRaimbault/CityNetwork/tree/master/Models/Synthetic}.

\subsubsection*{Results}

The study of the density model alone is developed in~\cite{raimbault2018calibration}. It is in particular calibrated on European density grid data, on 50km width square areas with 500m resolution for which real indicator values have been computed on whole Europe. Furthermore, a grid exploration of model behavior yields feasible output space in reasonable parameters bounds (roughly $\alpha \in [0.5,2],N_G\in [500,3000], P_m \in [10^4,10^5],\beta\in [0,0.2], n_d \in \{ 1, \ldots , 4\}$). The reduction of indicators space to a two dimensional plan through a Principal Component Analysis (variance explained with two components $\simeq 80\%$) allows to isolate a set of output points that covers reasonably precisely real point cloud. It confirms the ability of the model to reproduce morphologically the set of real configurations.

With a fixed population density, the conditional exploration of network generation model parameter space suggest a good flexibility on global indicators $\vec{G}$, together with good convergence properties. In order to apply the synthetic data generation method in relation with the thematical question of interactions between networks and territories, the exploration has been oriented towards the study of cross-correlations.

Given the large relative dimension of the parameter space, an exhaustive grid exploration is not possible. We use a Latin Hypercube sampling procedure with bounds given above for $\vec{\alpha}_D$ and for $\vec{\alpha}_N$, we take $N_C \in [50,120], r_g \in [1,100] , d_0 \in [0.1,10] , k_h \in [0,1] , \gamma \in [0.1,4],N_L\in [4,20]$. For number of model replications for each parameter point, less than 50 are enough to obtain confidence intervals at 95\% on indicators of width less than standard deviations. For correlations a hundred give confidence intervals (obtained with Fisher method) of size around 0.4, we take thus $n=80$ for experiments. The null model is simulated also with $n=80$, for random and density-based node distributions, and $r_o^{(0)} \in \{0.25,0.5,0.75\}$, $N_N^{(0)} \in \{10,15,20\}$ and $N_L^{(0)} \in \{20,30,40\}$. Simulation results are available on the dataverse at \url{http://dx.doi.org/10.7910/DVN/UIHBC7}.

\begin{figure}
\includegraphics[width=\linewidth]{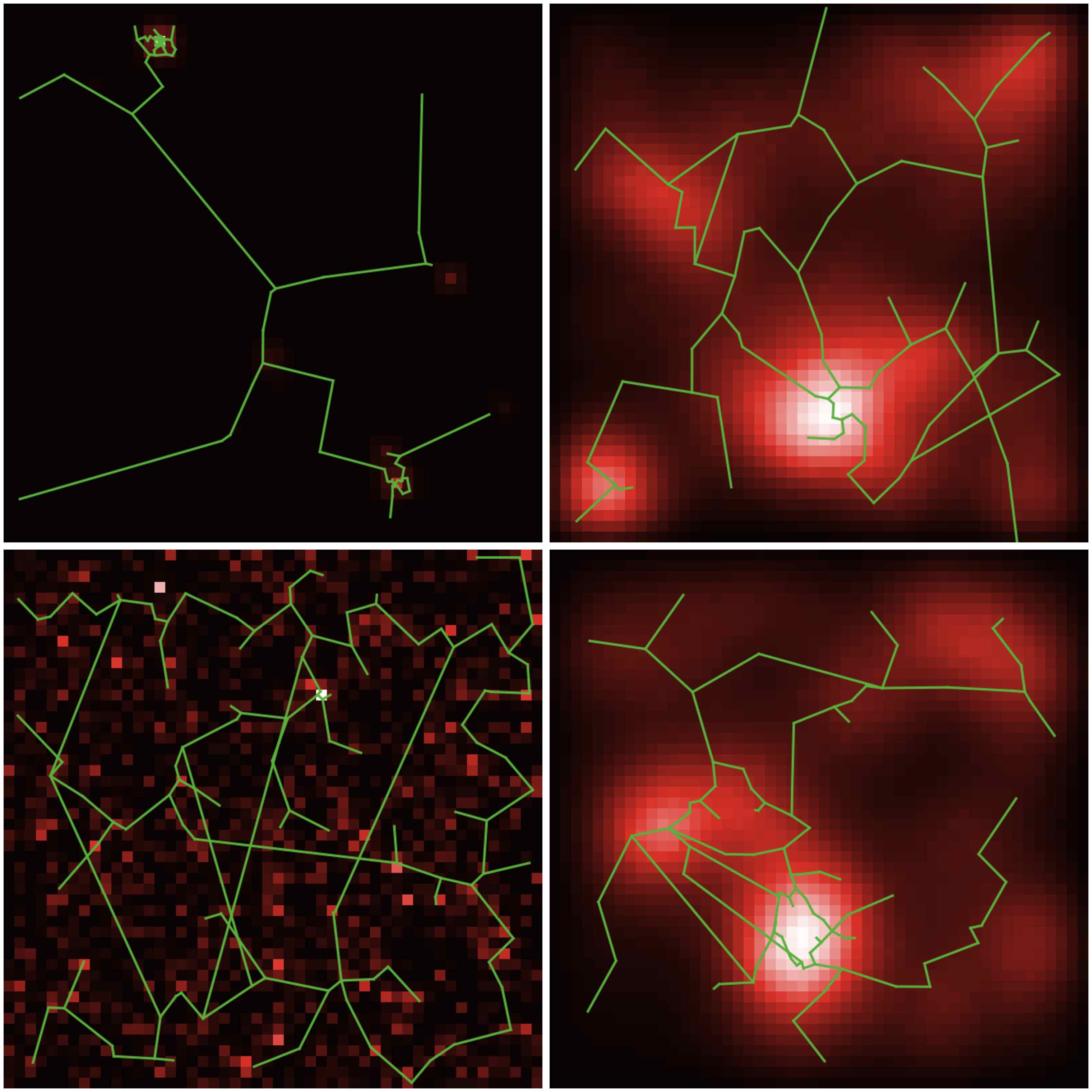}
	\caption{\textbf{Configurations obtained for parameters giving the four emphasized points in Fig.~\ref{fig:densnwcor}, in order from left to right and top to bottom.} We recognize polycentric city configurations (2 and 4), diffuse rural settlements (3) and aggregated weak density area (1). See appendice for exhaustive parameter values, indicators and corresponding correlations. For example $d$ is highly correlated with $l$ and $s$ ($\simeq$0.8) in (1) but not for (3) although both correspond to rural environments ; in the urban case we observe also a broad variability : $\rho[d,c]\simeq 0.34$ for (4) but $\simeq-0.41$ for (2), what is explained by a stronger role of gravitation hierarchy in (2) $\gamma=3.9,k_h=0.7$ (for (4), $\gamma=1.07,k_h=0.25$), whereas density parameters are similar.\label{fig:configexamples}}
\end{figure}

We show in Fig.~\ref{fig:configexamples} examples of generated territorial configurations. This visualization and some values of associated correlations already suggest that the method application yields a broad spectrum of generated correlation patterns. We obtain for example low density configurations, in aggregated or dispersed settings (top left, resp. bottom left panel), inducing very different types of networks. Similarly, areas with population centers which are closer to urban areas (Top right and bottom right panel), can also produce different network shapes. In the latest case, increasing the role of hierarchy through $\gamma$ and $k_h$ leads from a negative correlation between average distance $d$ and centrality $c$ to a positive correlation. This corresponds to a transition from processes where population dispersal decrease centrality (redundant networks) to inverse processes (centralized networks).

\begin{figure}
\includegraphics[width=\linewidth]{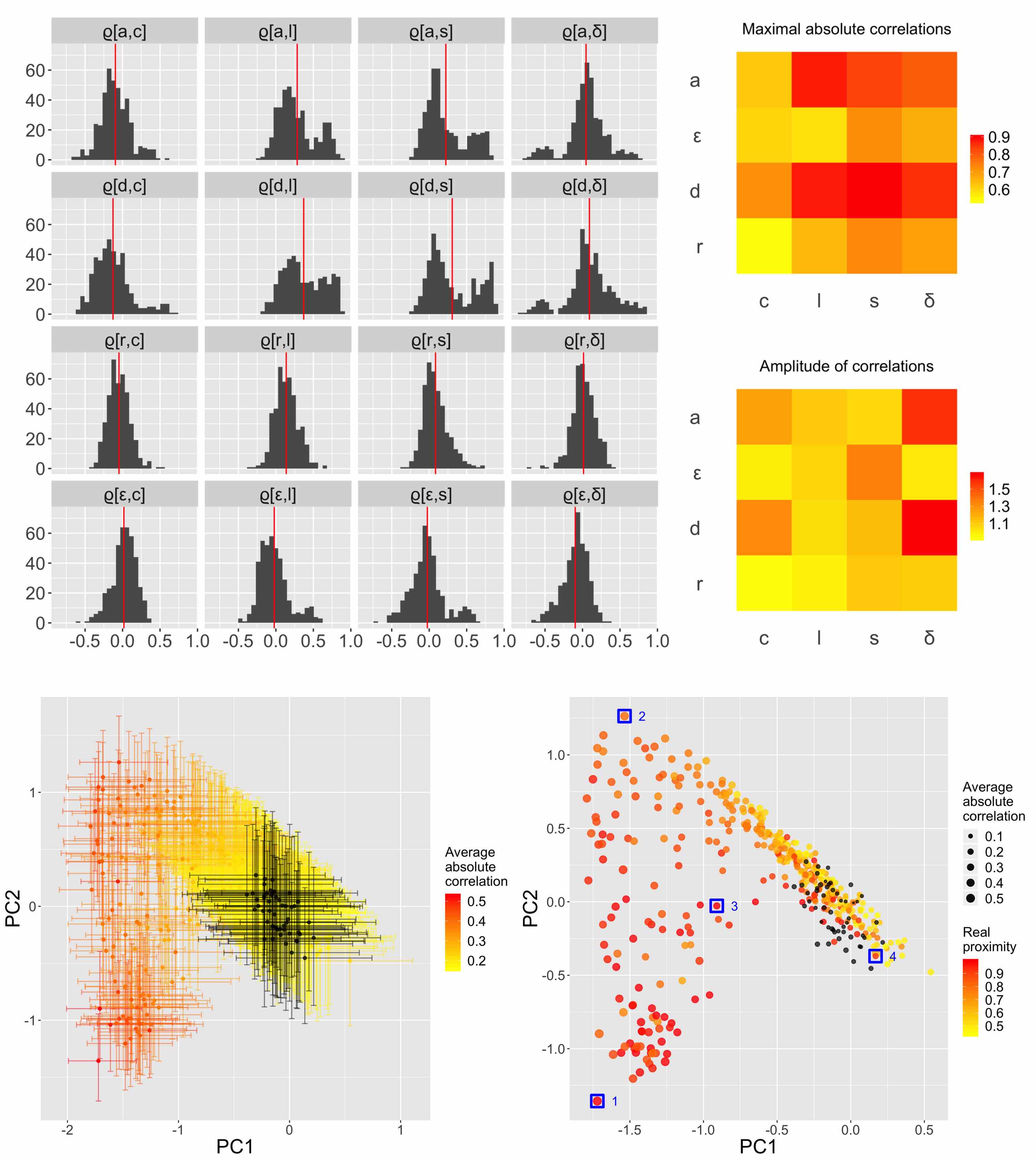}
\caption{\textbf{Exploration of feasible space for correlations between urban morphology and network structure.} \textbf{(Top left)} Statistical distribution of crossed-correlations between vectors $\vec{M}$ of morphological indicators (in numbering order Moran index, mean distance, entropy, hierarchy) and $\vec{N}$ of network measures (centrality, mean path length, speed, diameter). \textbf{(Top right)} Heatmaps for amplitude of correlations, defined as $a_{ij}=\max_k{\rho_{ij}^{(k)}}-\min_k{\rho_{ij}^{(k)}}$ and maximal absolute correlation, defined as $c_{ij}=\max_k\left| \rho_{ij}^{k} \right|$. \textbf{(Bottom left)} Projection of correlation matrices in a principal plan obtained by Principal Component Analysis on matrix population (cumulated variances: PC1=38\%, PC2=68\%). Error bars are initially computed as 95\% confidence intervals on each matrix element (by standard Fisher asymptotic method), and boundaries of confidence intervals are transformed into the component space. Scale color gives mean absolute correlation on full matrices. Black dots and error bars correspond to the realizations of the null model. \textbf{(Botton right)} Representation in the principal plan, scale color giving proximity to real data defined as $1 - \min_r \norm{\vec{M}-\vec{M}_r}$ where $\vec{M}_r$ is the set of real morphological measures, point size giving mean absolute correlation. The points highlighted in blue correspond to the configurations shown in Fig.~\ref{fig:configexamples}. Black dots correspond to the realizations of the null model.\label{fig:densnwcor}}
\end{figure}

Regarding the generation of correlated synthetic data in itself, several results presented in Fig.~\ref{fig:densnwcor} are worth noting. First of all, the statistical distributions of correlation coefficients (histograms, top left panel of Fig.~\ref{fig:densnwcor}) between morphology and network indicators are not systematically simple and some are bimodal. For example, the correlation $\rho[a,l]$ between hierarchy of the population distribution $a$ and mean path length in the network $l$ has a mode around 0 and an other around 0.6. This means that in a certain regime, these tend to decorrelate in average, while in an other regime they are strongly correlated. The latest correspond to configurations with a high Moran index and a high hierarchy, which means that more centralized urban configurations constrain the network path length through this correlation.

Second, still based on distributions in Fig.~\ref{fig:densnwcor}, but also on heatmaps for amplitude and maximal correlation (top right panel), we observe that it is possible to modulate up to a relatively high level of correlation for all indicators, since the maximal absolute correlation varies between 0.6 and 0.9. The amplitude of correlations ranges between 0.9 and 1.6, allowing thus a broad spectrum of values.

As the cross-correlation matrix is of dimension 16, we proceed to a principal component analysis on all generated correlation matrices (one matrix per row) to visualize the covered space in two dimensions. This PCA yields 38\% of variance for the first component and 68\% of cumulated variance for the second. We visualize the corresponding point cloud in the principal plan, with transformed confidence intervals (bottom left panel of Fig.~\ref{fig:densnwcor}) and with particular points (bottom right panel). The point cloud in the principal plan has a large extent but is not uniform: it is not possible to modulate in any direction any coefficient as they stay themselves correlated because of underlying generation processes. A more refined study at higher orders (correlation of correlations) would be necessary to precisely understand degrees of freedom in the generation of correlations. However, the covered area remains broad and confirms a rather flexible output space for generated correlations. When comparing to the null model runs (black dots and error bars), we find as expected that null model correlations are around 0 (all confidence intervals covering the origin), and that a part of the generated point cloud falls in the same area. An other important part of points fall outside the range of the null model in a statistically significant way. These are the interesting points for a possible application of the synthetic dataset, and we show thus that the method produces non-trivial and significant correlation patterns.

When evaluating the proximity of indicator values to real points (Equation~\ref{eq:data-proximity} in the abstract description of the method), which is given by the color level in the bottom right panel of Fig.~\ref{fig:densnwcor}, we note that the points with the highest level of correlation are also the ones which are closest to real data (red points). The points which are the farthest from real configurations are the uncorrelated ones, which also coincide with the null model. This suggests that in the frame of model hypotheses, real configurations exhibit high correlations between network properties and urban form. \cite{raimbault2019urban} confirms this fact by studying real effective correlations.

Finally, some examples of configurations taken on particular points in the principal plan, highlighted in blue in Fig.~\ref{fig:densnwcor} and described above (Fig.~\ref{fig:configexamples}), show that similar population density profiles can yield very different correlation profiles. This confirms the flexibility of the method and the possibility to control on correlation structure.

\section*{Correlated financial time-series}

We also apply the method to a totally different type of system, namely financial complex systems. Financial time-series are heterogeneous, multi-scalar and non-stationary~\cite{mantegna2000introduction}. Correlations are broadly explored in that field. For example, Random Matrix Theory allows to distinguish signal from noise for a correlation matrix computed for a large number of asset with low-frequency signals, typically with a time scale of a day \cite{2009arXiv0910.1205B}. Similarly, Complex Network Analysis on networks constructed from correlations introduced methods such as Minimal Spanning Tree~\cite{2001PhyA..299...16B} or more refined topologically constrained network generation methods~\cite{tumminello2005tool}. These provide reconstructions of economic sectors structure. At high frequencies, the precise estimation of interdependence parameters assuming models for asset dynamics has been extensively studied from a theoretical point of view aimed at refinement of models and estimators~\cite{barndorff2011multivariate}. Theoretical results must be tested on synthetic datasets as they ensure a control of most parameters in order to check that a predicted effect is indeed observable all things being otherwise equal. Empirical confirmation of the improvement of estimators is obtained on a synthetic dataset at a fixed correlation level.

We consider a network of assets $(X_i(t))_{1\leq i \leq N}$ sampled at high-frequency (typically 1s). We use a multi-scalar framework (used e.g. in wavelet analysis approaches~\cite{ramsey2002wavelets} or in multi-fractal signal processing~\cite{bouchaud2000apparent}) to interpret observed signals as the superposition of components at different time scales. We thus write $X_i=\sum_{\omega}{X_i^{\omega}}$. We denote by $T_i^{\omega} = \sum_{\omega' \leq \omega} X_i^{\omega}$ the filtered signal at a given frequency $\omega$. A typical problem in the study of complex systems is the prediction of a trend at a given scale. It can be viewed as the identification of regularities and their distinction from components considered as random. For the sake of simplicity, we represent such a process as a trend prediction model at a given temporal scale $\omega_1$, formally an estimator $M_{\omega_1} : (T_i^{\omega_1}(t'))_{t'<t} \mapsto \hat{T_i}^{\omega_1}(t)$ which aims to minimize error on the real trend $\norm{T_i^{\omega_1} - \hat{T}_i^{\omega_1}}$. In the case of autoregressive multivariate estimators, the performance will depend among other parameters on respective correlations between assets. It is thus interesting to apply the method to the evaluation of performance as a function of correlation at different scales. We assume a Black-Scholes dynamic for assets~\cite{jarrow1999honor}, i.e. $dX = \sigma\cdot dW$, with $W$ Wiener process. Such a dynamic model allows an easy modulation of correlation levels.

\begin{figure}
\includegraphics[width=\columnwidth]{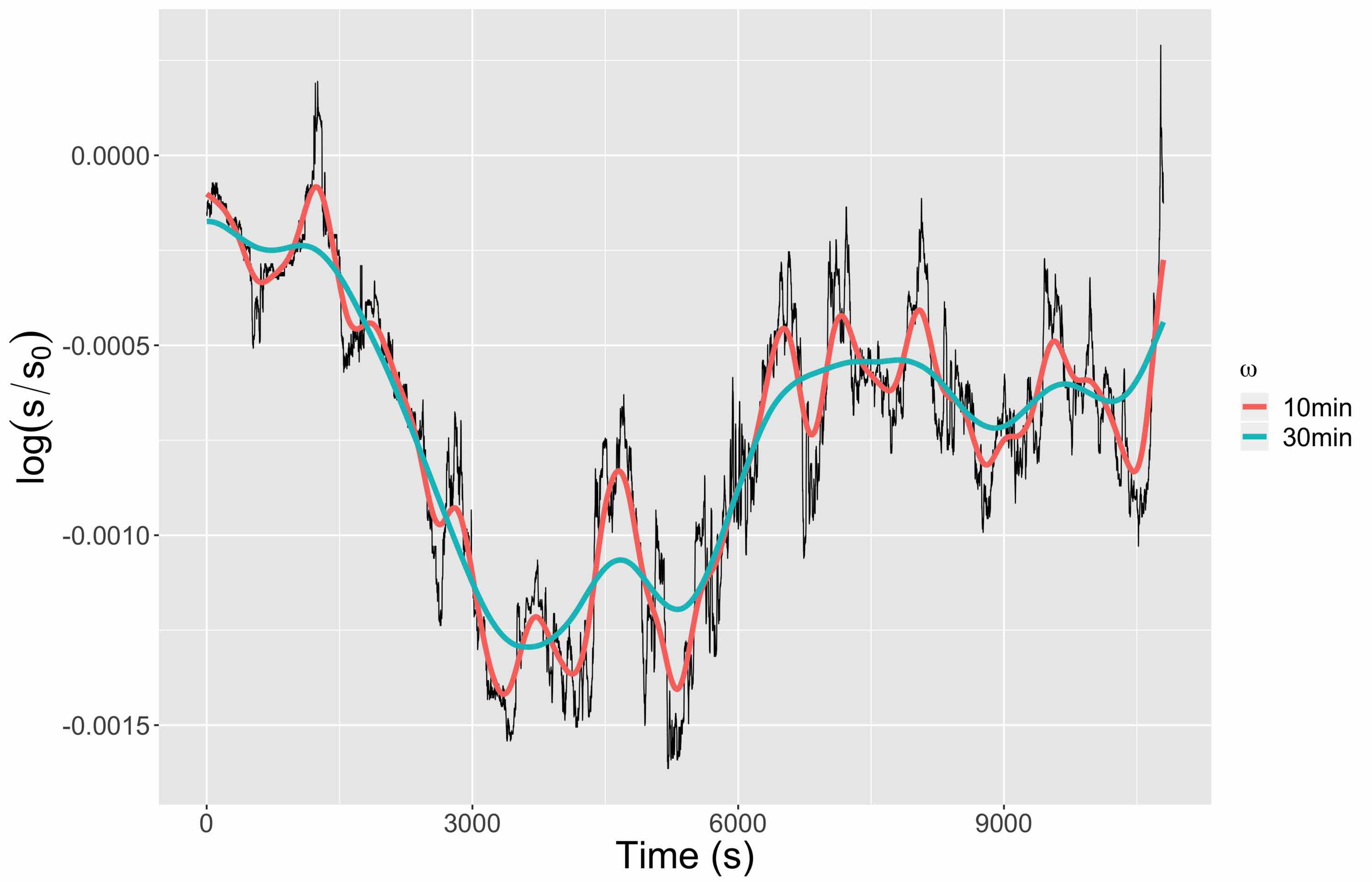}
\caption{\textbf{Example of the multi-scalar structure of the signal, basis of the construction of synthetic signals.} \emph{Log-prices} are represented on a time window of around 3h for November 1st 2015 for asset EUR/USD, with 10min (purple) and 30min trends. The low-frequency components are the basis to build synthetic data, on which noises with a controlled correlation are superposed.\label{fig:example_signal}}
\end{figure}

\subsection*{Data generation}

We can straightforward generate $\tilde{X}_i$ such that $\Varb{\tilde{X}_i^{\omega_1}}= ^{t}\mathbf{\Sigma} \cdot \mathbf{R} \cdot \mathbf{\Sigma}$ (with $\mathbf{\Sigma}$ are estimated standard deviations and $\mathbf{R}$ is a fixed correlation matrix) and verifying $X_i^{\omega \leq \omega_0} = \tilde{X}_i^{\omega \leq \omega_0}$. This means in practice that the data proximity indicator is the identity of components at a lower frequency than a fundamental frequency $\omega_0 < \omega_1$. We use therefore the simulation of Wiener processes with fixed correlation. Indeed, if $dW_1 \indep dW_1^{\indep}$ (and $\sigma_1 < \sigma_2$ indicatively, assets being interchangeable), then
\begin{equation}
\label{eq:orthogonal}
W_2 = \rho_{12}W_1 + \sqrt{1-\frac{\sigma_1^2}{\sigma_2^2}\cdot\rho_{12}^2}\cdot W_1^{\indep}
\end{equation}
is such that $\rho(dW_1,dW_2)=\rho_{12}$. Signals for other components can be constructed the same way by Gram orthonormalization. We isolate the component at the desired frequency $\omega_1$ by filtering the signal, i.e. using signals constructed with Eq.~\ref{eq:orthogonal} such that $\tilde{X}_i^{\omega_1} = W_i - \mathcal{F}_{\omega_0}[W_i]$, where $\mathcal{F}_{\omega_0}$ is a low-pass filter with cut-off frequency $\omega_0$. We reconstruct then the hybrid synthetic signals by taking
 
\begin{equation}
\tilde{X}_i = T_i^{\omega_0} + \tilde{X}_i^{\omega_1}
\end{equation}

The method is tested on an example with two assets from foreign exchange market (EUR/USD and EUR/GBP), on a six month period from June 2015 to November 2015. Data was obtained from \url{http://www.histdata.com/}. The data cleaning procedure, starting from original series sampled at a frequency around 1s, consists in a first step to the determination of the minimal common temporal range (missing sequences being ignored, by vertical translation of series, i.e. $S(t):=S(t)\cdot \frac{S(t_{n})}{S(t_{n-1})}$ when $t_{n-1},t_n$ are extremities of the ``hole'' and $S(t)$ value of the asset, what is equivalent to keep the constraint to have returns at similar temporal steps between assets). We study then \emph{log-prices} and \emph{log-returns} \cite{mantegna2000introduction}, defined by $X(t):=\log{\frac{S(t)}{S_0}}$ and $\Delta X (t) = X(t) - X(t-1)$. Raw data are filtered at a maximal frequency $\omega_m = 10\textrm{min}$ (which will be the maximal frequency for following treatments) for concerns of computational efficiency. As time-series are then sampled at $3\cdot\omega_m$ to avoid aliasing, a day of size 86400 for 1s sampling is reduced to a much smaller size of 432. We use a non-causal gaussian filter of total width $\omega$. We fix the fundamental frequency $\omega_0=24\textrm{h}$ and we propose to construct synthetic data at frequencies $\omega_1 = 30\textrm{min},1\textrm{h},2\textrm{h}$. We show in Fig.~\ref{fig:example_signal} an example of signal structure at the scales $\omega_m$ and $\omega_1 = 30\textrm{min}$, compared with the non-filtered raw signal.

It is crucial to consider the interference between $\omega_0$ and $\omega_1$ frequencies in the reconstructed signal: the correlation which is indeed estimated is 
\begin{equation}
\rho_{e} = \rho \left[ \Delta \tilde{X}_1 , \Delta \tilde{X}_2 \right] = \rho \left[ \Delta T_1^{\omega_0} + \Delta \tilde{X}_1^{\omega} , \Delta T_2^{\omega_0} + \Delta \tilde{X}_2^{\omega}\right]
\end{equation}

Assuming to be in the reasonable limit $\sigma_1 \gg \sigma_0$ (fundamental frequency small enough), that $\Covb{\Delta \tilde{X}_i^{\omega_1}}{\Delta X_j^{\omega}}=0$ for all $i,j,\omega_1 > \omega$ and that returns are centered at any scale, we can develop the previous expression to compute the correction on effective correlation due to interferences. We obtain at the first order the expression of effective correlation given by

\begin{equation}
\label{eq:eff_corr}
\rho_e = \left[ \varepsilon_1 \varepsilon_2 \rho_0 + \rho \right] \cdot \left[ 1 - \frac{1}{2}\left(\varepsilon_1^2 + \varepsilon_2^2 \right) \right]
\end{equation}

{\noindent}what corresponds to the correlation that we can effectively simulate in synthetic data.

Correlations are in practice estimated with a Pearson estimator, the covariances being corrected for bias, i.e. following Eq.~\ref{eq:correstimate}, Eq.~\ref{eq:covariance} and Eq.~\ref{eq:variance}.

The generated synthetic data are then used to test a toy model. We propose in particular to investigate the predictive power of a very simple linear model. The tested predictive model $M_{\omega_1}$ is a simple \emph{ARMA} for which parameters $p=2,q=0$ are fixed (as we do not create lagged correlation, we do not expect large orders of auto-regression as these kind of processes have short memory for real data; furthermore smoothing is not necessary as data are already filtered). It is however applied in an adaptive way, in the sense that given a time window $T_W$, we estimate for any $t$ the model on $[t-T_W+1,t]$ in order to predict signals at $t+1$.

Experiments are implemented in the \texttt{R} language, using in particular the \texttt{MTS}~\cite{Tsay:2015xy} library for time-series models. Cleaned data and source code are available on an open \texttt{git} repository at \url{https://github.com/JusteRaimbault/SyntheticAsset}.

\begin{figure}
\centering
\includegraphics[width=\linewidth]{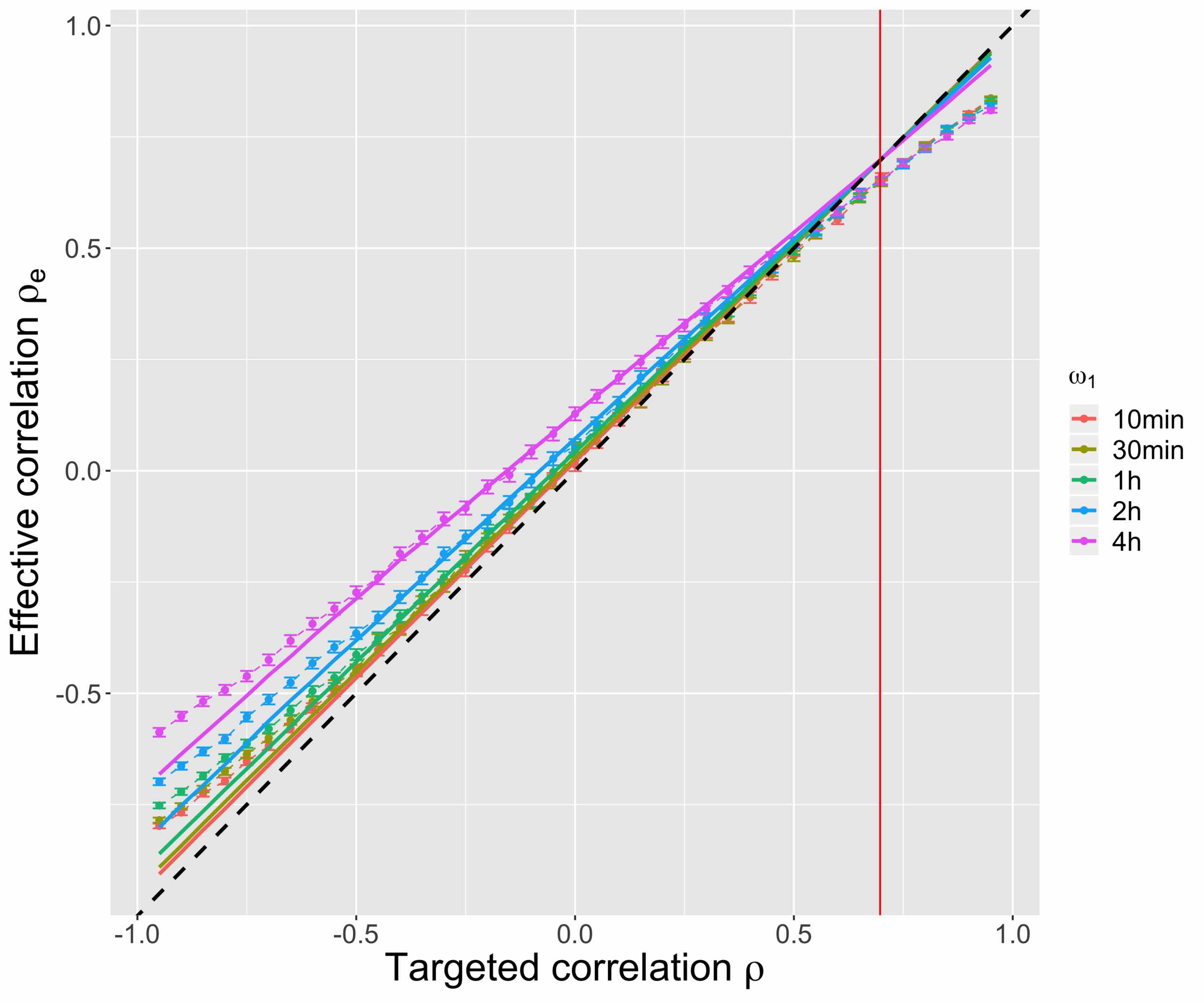}
\caption{\textbf{Effective correlations obtained on synthetic data.} Dots represent estimated correlations on a synthetic dataset corresponding to 6 months between June and November 2015 (error-bars give 95\% confidence intervals obtained with standard Fisher method); scale color gives the filtering frequency $\omega_1=10\textrm{min},30\textrm{min},1\textrm{h},2\textrm{h},4\textrm{h}$; solid lines give the theoretical values for $\rho_e$ obtained by~\ref{eq:eff_corr} with estimated volatilities (dotted-line diagonal for reference); vertical red line position is the theoretical value such that $\rho = \rho_e$ with mean values for $\varepsilon_i$ on all points. We observe for high absolute correlations values a deviation from corrected values, what should be caused by non-verified independence and centered returns assumptions. Asymmetry is caused by the high value of $\rho_0 \simeq 0.71$.}
\label{fig:effective_corrs}
\end{figure}

\begin{figure}
\includegraphics[width=\linewidth]{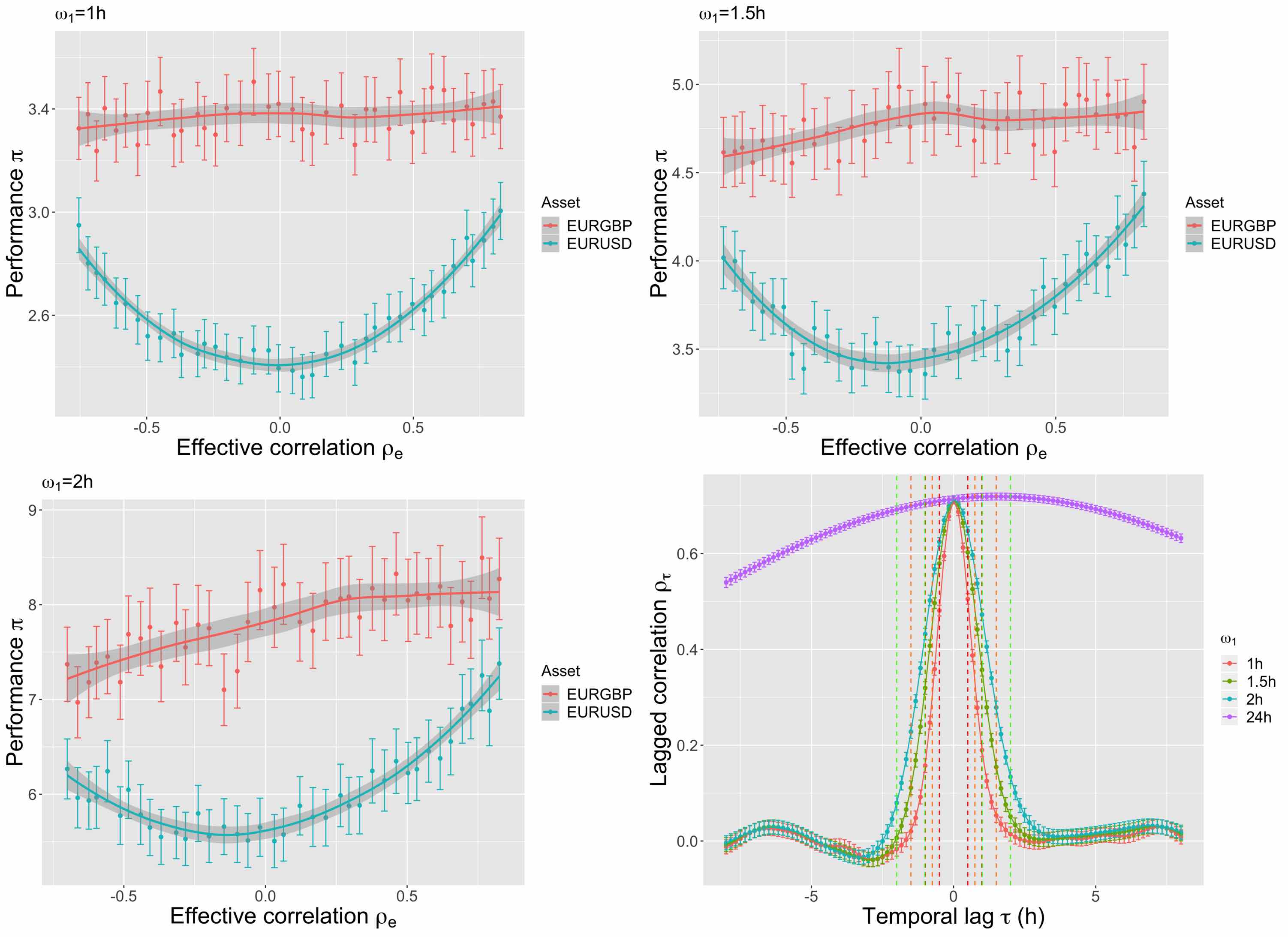}
\caption{\textbf{Performance of a predictive model as a function of simulated correlations.} From left to right and top to bottom, the plots show for each asset the normalized performance of an ARMA model ($p=2,q=0$), defined as $\pi = \left(\frac{1}{T}\sum_t\left(\tilde{X}_i(t) - M_{\omega_1}\left[\tilde{X}_i\right](t)\right)^2 \right) / \sigma \left[ \tilde{X}_i \right]^2$ (95\% confidence intervals computed by $\pi = \bar{\pi} \pm (1.96\cdot \sigma [\pi])/\sqrt{T}$, local polynomial smoothing to ease reading). It is interesting to note the U-shape for EUR/USD, due to interference between components at different scales. Correlation between simulated noises deteriorates predictive power. The study of lagged correlations (here $\rho [\Delta X_{\textrm{EURUSD}}(t),\Delta X_{\textrm{EURGBP}}(t-\tau)]$) on real data clarifies this phenomenon: the fourth graph show an asymmetry in curves at any scale compared to zero lag $(\tau = 0)$ what leads fundamental components to increase predictive power for the dollar, amelioration then perturbed by correlations between simulated components. Dashed lines show time steps (in equivalent $\tau$ units) used by the ARMA at each scale, what allows to read the corresponding lagged correlation on fundamental component.\label{fig:model_perf}}
\end{figure}

Fig.~\ref{fig:effective_corrs} shows the effective correlations computed on synthetic data. For standard parameter values (for example $\omega_0=24\textrm{h}$, $\omega_1=2\textrm{h}$ and $\rho=-0.5$), we find $\rho_0\simeq 0.71$ et $\varepsilon_i \simeq 0.3$ what yields $\left| \rho_e - \rho \right|\simeq 0.05$. We observe a good agreement between observed $\rho_e$ and values predicted by Equation~\ref{eq:eff_corr} in the interval $\rho \in [-0.5,0.5]$. On the contrary, for larger absolute values, a deviation increasing with $\left|\rho\right|$ and as $\omega_1$ decreases : it confirms the intuition that when frequency decreases and becomes closer to $\omega_0$, interferences between the two components are not negligible anymore and invalidate independence assumptions for example.

\subsection*{Application to the study of a predictive model performance}

The predictive model described above is then applied to synthetic data, in order to study its average performance as a function of correlation between signals. Results for $\omega_1 = 1\textrm{h},1\textrm{h}30,2\textrm{h}$ are shown in Fig.~\ref{fig:model_perf}. The a priori counter-intuitive result of a maximal performance at vanishing correlation for one of the assets confirms the role of synthetic data to better understand system mechanisms : the study of lagged correlations shows an asymmetry in the real data that we can understand at a daily scale as an increased influence of EUR/GBP on EUR/USD with a rough two hours lag. The existence of this \emph{lag} allows a ``good'' prediction of EUR/USD thanks to fundamental component. This predictive power is perturbed by added noises in a way that increases with their correlation. The more noises correlated are, the more he model will take them into account and will make false predictions because of the Markovian character of simulated brownian (note that the model used has theoretically no predictive power at all on pure brownians).

This case study on a \emph{toy-model} illustrates the relevance of using simulated synthetic data. Further developments can be directed towards the simulation of more realistic data (presence of consistent \emph{lagged correlation} patterns, more realistic models than Black-Scholes) and apply it on more operational predictive models.

\section*{Discussion}

\subsection*{Contributions}

We investigated in this paper the possibility of generating synthetic data at a macrosopic level with a controlled correlation structure. The generic method we introduce can be applied to any complex system, where the proximity to real data is measured on aggregated indicators. The method was designed more particularly for socio-spatial systems. We show in the case of transportation network and territories, by exploring a weak coupled model for population density and road network generation, that varying model parameters yield a broad output space of effective correlations. Two configurations with the same first order indicator values can capture very different underlying correlations. This means that future applications to the study of upstream models to the sensitivity of spatial initial configuration, such as the one done by \cite{raimbault2018space} but in which correlation structure is controlled, should be made possible by our approach.

We postulate that the method can be also applied in other fields where similar constraints can be of interest. Indeed, in the context of financial data, considering data proximity indicators based on low-frequency components of signals, we showed how correlation can be controlled and even analytically predicted to a certain extent. Our work recalls thus the interest in generating hybrid data, and is differentiated from most work where only the microscopic level is taken into account.

As already mentioned, most of simulation models need an initial state generated artificially as soon as model parametrization is not done completely on real data. An advanced model sensitivity analysis implies a control on parameters for synthetic dataset generation, seen as model meta-parameters~\cite{raimbault2018space}. In the case of a statistical analysis of model outputs it provides a way to operate a second order statistical control.

\subsection*{Future work}

Regarding the application to geographical data, the calibration of the network generation component at given density, on real data for transportation network, is a potential development. It would be relevant typically on road networks given the shape of generated networks, what should be possible using OpenStreetMap open data which have a reasonable quality for Europe~\cite{girres2010quality}. This should theoretically allow to unveil parameter sets reproducing accurately existing configurations both for urban morphology and network shape. By attributing a synthetic dataset similar to a given real configuration, we would be able to compute a sort of \emph{intrinsic correlation} proper to this configuration.

We studied in the second example stochastic processes in the sense of random time-series, whereas time did not have a role in the first case. We can suggest a strong coupling between the two model components (or the construction of an integrated model) and to observe indicators and correlations at different time steps during the generation. In dynamical spatial models the existence of lagged interdependences in space and time~\cite{pigozzi1980interurban} is an important feature of complex dynamics. This would provide a better understanding of the link between static and dynamic correlations.

We were limited to the control of first and second moments of generated data, but we could imagine a theoretical generalization allowing the control of moments at any order. However, as shown by the geographical example, the difficulty of generation in a concrete complex case questions the possibility of higher orders control when keeping a consistent structure model and a reasonable number of parameters. The study of non-linear dependence structures as proposed in~\cite{chicheportiche2013nested} is in an other perspective an interesting possible development.

We could also apply specific exploration algorithms to explore more exhaustively the feasible correlation space. Such an algorithm based on Novelty Search has been introduced by~\cite{10.1371/journal.pone.0138212}. Coupling it with our method would allow establishing the full range of feasible correlations for a given generation model.

\section*{Conclusion}

We proposed an abstract method to generate synthetic datasets in which correlation structure is controlled, but the empirical data required can be sparse or targeting macroscopic aggregated criteria. Its implementation in two very different fields shows its flexibility and the broad range of possible applications. More generally, it is crucial to favorise such practices of systematic validation of computational models by statistical analysis, in particular for agent-based models for which the question of validation remains an open issue.

Furthermore, our overall approach enters a particular epistemological frame. On the one hand it has a strong multidisciplinary aspect, and on the other hand the importance of empirical component through computational exploration methods make this approach typical of Complex Systems science \cite{2009arXiv0907.2221B}. The combination of empirical knowledge obtained from data mining, with knowledge obtained by modeling and simulation is generally central to the conception and exploration of multi-scalar heterogeneous models. The method and results presented here is an illustration of such an hybrid paradigm.

\section*{Availability of data and material}

All data and code used in this study, including simulation results, are openly available on git repositories at \url{https://github.com/JusteRaimbault/CityNetwork/tree/master/Models/Synthetic} and \url{https://github.com/JusteRaimbault/SyntheticAsset}. Large dataset are available on the dataverse repository at \url{http://dx.doi.org/10.7910/DVN/UIHBC7}.

\vspace{-0.2cm}

\section*{Competing interests}

The author declares to have no competing interests.

\vspace{-0.2cm}

\section*{Funding}

This work is part of DynamiCity, a FUI project funded by BPI France, Auvergne-Rh{\^o}ne-Alpes region, Ile-de-France region and Lyon metropolis. This work was also funded by the Urban Dynamics Lab grant EPSRC EP/M023583/1.

\vspace{-0.2cm}

\section*{Authors' contributions}

JR designed the study, did the analysis and wrote the paper.

\vspace{-0.2cm}

\section*{Acknowledgements}

Results obtained in this paper were computed on the vo.complex-system.eu virtual organization of the European Grid Infrastructure ( http://www.egi.eu ). We thank the European Grid Infrastructure and its supporting National Grid Initiatives (France-Grilles in particular) for providing the technical support and infrastructure. The author thanks the organizers and participants of Journ{\'e}es de Rochebrune 2016 for which the work was originally conceived. The author also thanks E. Marandon (L2 Technologies) for the original idea of correlated financial signals.

 \vspace{-0.2cm}
  

\newcommand{\BMCxmlcomment}[1]{}

\BMCxmlcomment{

<refgrp>

<bibl id="B1">
  <title><p>Synthetic control methods for comparative case studies: Estimating
  the effect of California's tobacco control program</p></title>
  <aug>
    <au><snm>Abadie</snm><fnm>A</fnm></au>
    <au><snm>Diamond</snm><fnm>A</fnm></au>
    <au><snm>Hainmueller</snm><fnm>J</fnm></au>
  </aug>
  <source>Journal of the American Statistical Association</source>
  <pubdate>2010</pubdate>
  <volume>105</volume>
  <issue>490</issue>
</bibl>

<bibl id="B2">
  <title><p>Creating a synthetic population</p></title>
  <aug>
    <au><snm>Moeckel</snm><fnm>R</fnm></au>
    <au><snm>Spiekermann</snm><fnm>K</fnm></au>
    <au><snm>Wegener</snm><fnm>M</fnm></au>
  </aug>
  <source>Proceedings of the 8th International Conference on Computers in Urban
  Planning and Urban Management (CUPUM)</source>
  <pubdate>2003</pubdate>
</bibl>

<bibl id="B3">
  <title><p>Advances in agent population synthesis and application in an
  integrated land use and transportation model</p></title>
  <aug>
    <au><snm>Pritchard</snm><fnm>DR</fnm></au>
    <au><snm>Miller</snm><fnm>EJ</fnm></au>
  </aug>
  <source>Transportation Research Board 88th Annual Meeting</source>
  <pubdate>2009</pubdate>
  <issue>09-1686</issue>
</bibl>

<bibl id="B4">
  <title><p>A review of feature selection methods on synthetic data</p></title>
  <aug>
    <au><snm>Bol{\'o}n Canedo</snm><fnm>V</fnm></au>
    <au><snm>S{\'a}nchez Maro{\~n}o</snm><fnm>N</fnm></au>
    <au><snm>Alonso Betanzos</snm><fnm>A</fnm></au>
  </aug>
  <source>Knowledge and information systems</source>
  <publisher>Springer</publisher>
  <pubdate>2013</pubdate>
  <volume>34</volume>
  <issue>3</issue>
  <fpage>483</fpage>
  <lpage>-519</lpage>
</bibl>

<bibl id="B5">
  <title><p>SynTReN: a generator of synthetic gene expression data for design
  and analysis of structure learning algorithms</p></title>
  <aug>
    <au><snm>Bulcke</snm><fnm>T</fnm></au>
    <au><snm>Van Leemput</snm><fnm>K</fnm></au>
    <au><snm>Naudts</snm><fnm>B</fnm></au>
    <au><snm>Remortel</snm><fnm>P</fnm></au>
    <au><snm>Ma</snm><fnm>H</fnm></au>
    <au><snm>Verschoren</snm><fnm>A</fnm></au>
    <au><snm>De Moor</snm><fnm>B</fnm></au>
    <au><snm>Marchal</snm><fnm>K</fnm></au>
  </aug>
  <source>BMC bioinformatics</source>
  <publisher>BioMed Central Ltd</publisher>
  <pubdate>2006</pubdate>
  <volume>7</volume>
  <issue>1</issue>
  <fpage>43</fpage>
</bibl>

<bibl id="B6">
  <title><p>Creating synthetic baseline populations</p></title>
  <aug>
    <au><snm>Beckman</snm><fnm>RJ</fnm></au>
    <au><snm>Baggerly</snm><fnm>KA</fnm></au>
    <au><snm>McKay</snm><fnm>MD</fnm></au>
  </aug>
  <source>Transportation Research Part A: Policy and Practice</source>
  <publisher>Elsevier</publisher>
  <pubdate>1996</pubdate>
  <volume>30</volume>
  <issue>6</issue>
  <fpage>415</fpage>
  <lpage>-429</lpage>
</bibl>

<bibl id="B7">
  <title><p>Population synthesis for microsimulation: State of the
  art</p></title>
  <aug>
    <au><snm>M{\"u}ller</snm><fnm>K</fnm></au>
    <au><snm>Axhausen</snm><fnm>KW</fnm></au>
  </aug>
  <source>Arbeitsberichte Verkehrs-und Raumplanung</source>
  <publisher>IVT, ETH Zurich</publisher>
  <pubdate>2010</pubdate>
  <volume>638</volume>
</bibl>

<bibl id="B8">
  <title><p>Synthetic population generation without a sample</p></title>
  <aug>
    <au><snm>Barthelemy</snm><fnm>J</fnm></au>
    <au><snm>Toint</snm><fnm>PL</fnm></au>
  </aug>
  <source>Transportation Science</source>
  <publisher>INFORMS</publisher>
  <pubdate>2013</pubdate>
  <volume>47</volume>
  <issue>2</issue>
  <fpage>266</fpage>
  <lpage>-279</lpage>
</bibl>

<bibl id="B9">
  <title><p>Synthetic data generation: Theory, techniques and
  applications</p></title>
  <aug>
    <au><snm>Hoag</snm><fnm>JE</fnm></au>
  </aug>
  <publisher>Ann Arbor: University of Arkansas</publisher>
  <pubdate>2008</pubdate>
</bibl>

<bibl id="B10">
  <title><p>Generating synthetic data to match data mining patterns</p></title>
  <aug>
    <au><snm>Eno</snm><fnm>J</fnm></au>
    <au><snm>Thompson</snm><fnm>CW</fnm></au>
  </aug>
  <source>IEEE Internet Computing</source>
  <publisher>IEEE</publisher>
  <pubdate>2008</pubdate>
  <volume>12</volume>
  <issue>3</issue>
  <fpage>78</fpage>
  <lpage>-82</lpage>
</bibl>

<bibl id="B11">
  <title><p>Complexity and the Shift in Modern Science</p></title>
  <aug>
    <au><snm>Arthur</snm><fnm>WB</fnm></au>
  </aug>
  <source>Conference on Complex Systems, Tempe, Arizona</source>
  <pubdate>2015</pubdate>
</bibl>

<bibl id="B12">
  <title><p>Investigation of Underlying Distributional Assumption in Nested
  Logit Model Using Copula-Based Simulation and Numerical
  Approximation</p></title>
  <aug>
    <au><snm>Ye</snm><fnm>X</fnm></au>
  </aug>
  <source>Transportation Research Record: Journal of the Transportation
  Research Board</source>
  <publisher>Transportation Research Board of the National
  Academies</publisher>
  <pubdate>2011</pubdate>
  <issue>2254</issue>
  <fpage>36</fpage>
  <lpage>-43</lpage>
</bibl>

<bibl id="B13">
  <title><p>SYNTHESIS—a synthetic spatial information system for urban and
  regional analysis: methods and examples</p></title>
  <aug>
    <au><snm>Birkin</snm><fnm>M</fnm></au>
    <au><snm>Clarke</snm><fnm>M</fnm></au>
  </aug>
  <source>Environment and planning A</source>
  <publisher>SAGE Publications Sage UK: London, England</publisher>
  <pubdate>1988</pubdate>
  <volume>20</volume>
  <issue>12</issue>
  <fpage>1645</fpage>
  <lpage>-1671</lpage>
</bibl>

<bibl id="B14">
  <title><p>Differentially private synthesization of multi-dimensional data
  using copula functions</p></title>
  <aug>
    <au><snm>Li</snm><fnm>H</fnm></au>
    <au><snm>Xiong</snm><fnm>L</fnm></au>
    <au><snm>Jiang</snm><fnm>X</fnm></au>
  </aug>
  <source>Advances in database technology: proceedings. International
  Conference on Extending Database Technology</source>
  <pubdate>2014</pubdate>
  <volume>2014</volume>
  <fpage>475</fpage>
</bibl>

<bibl id="B15">
  <title><p>The structure and function of complex networks</p></title>
  <aug>
    <au><snm>Newman</snm><fnm>ME</fnm></au>
  </aug>
  <source>SIAM review</source>
  <publisher>SIAM</publisher>
  <pubdate>2003</pubdate>
  <volume>45</volume>
  <issue>2</issue>
  <fpage>167</fpage>
  <lpage>-256</lpage>
</bibl>

<bibl id="B16">
  <title><p>Cross-correlated random field generation with the direct Fourier
  transform method</p></title>
  <aug>
    <au><snm>Robin</snm><fnm>MJL</fnm></au>
    <au><snm>Gutjahr</snm><fnm>AL</fnm></au>
    <au><snm>Sudicky</snm><fnm>EA</fnm></au>
    <au><snm>Wilson</snm><fnm>JL</fnm></au>
  </aug>
  <source>Water Resources Research</source>
  <publisher>Wiley Online Library</publisher>
  <pubdate>1993</pubdate>
  <volume>29</volume>
  <issue>7</issue>
  <fpage>2385</fpage>
  <lpage>-2397</lpage>
</bibl>

<bibl id="B17">
  <title><p>A multilevel, hierarchical sampling technique for spatially
  correlated random fields</p></title>
  <aug>
    <au><snm>Osborn</snm><fnm>S</fnm></au>
    <au><snm>Vassilevski</snm><fnm>PS</fnm></au>
    <au><snm>Villa</snm><fnm>U</fnm></au>
  </aug>
  <source>SIAM Journal on Scientific Computing</source>
  <publisher>SIAM</publisher>
  <pubdate>2017</pubdate>
  <volume>39</volume>
  <issue>5</issue>
  <fpage>S543</fpage>
  <lpage>-S562</lpage>
</bibl>

<bibl id="B18">
  <title><p>Regional-scale geostatistical inverse modeling of North American CO
  2 fluxes: a synthetic data study</p></title>
  <aug>
    <au><snm>Gourdji</snm><fnm>SM</fnm></au>
    <au><snm>Hirsch</snm><fnm>AI</fnm></au>
    <au><snm>Mueller</snm><fnm>KL</fnm></au>
    <au><snm>Yadav</snm><fnm>V</fnm></au>
    <au><snm>Andrews</snm><fnm>AE</fnm></au>
    <au><snm>Michalak</snm><fnm>AM</fnm></au>
  </aug>
  <source>Atmospheric Chemistry and Physics</source>
  <publisher>Copernicus GmbH</publisher>
  <pubdate>2010</pubdate>
  <volume>10</volume>
  <issue>13</issue>
  <fpage>6151</fpage>
  <lpage>-6167</lpage>
</bibl>

<bibl id="B19">
  <title><p>Simulation of cross-correlated random field samples from sparse
  measurements using Bayesian compressive sensing</p></title>
  <aug>
    <au><snm>Zhao</snm><fnm>T</fnm></au>
    <au><snm>Wang</snm><fnm>Y</fnm></au>
  </aug>
  <source>Mechanical Systems and Signal Processing</source>
  <publisher>Elsevier</publisher>
  <pubdate>2018</pubdate>
  <volume>112</volume>
  <fpage>384</fpage>
  <lpage>-400</lpage>
</bibl>

<bibl id="B20">
  <title><p>Geosimulation: Automata-based modeling of urban
  phenomena</p></title>
  <aug>
    <au><snm>Benenson</snm><fnm>I</fnm></au>
    <au><snm>Torrens</snm><fnm>P</fnm></au>
  </aug>
  <publisher>Chichester: John Wiley \& Sons</publisher>
  <pubdate>2004</pubdate>
</bibl>

<bibl id="B21">
  <title><p>The new science of cities</p></title>
  <aug>
    <au><snm>Batty</snm><fnm>M</fnm></au>
  </aug>
  <publisher>Cambridge: MIT Press</publisher>
  <pubdate>2013</pubdate>
</bibl>

<bibl id="B22">
  <title><p>An Evolutionary Theory of Urban Systems</p></title>
  <aug>
    <au><snm>Pumain</snm><fnm>D</fnm></au>
  </aug>
  <source>International and Transnational Perspectives on Urban
  Systems</source>
  <publisher>Singapore: Springer</publisher>
  <pubdate>2018</pubdate>
  <fpage>3</fpage>
  <lpage>-18</lpage>
</bibl>

<bibl id="B23">
  <title><p>Simulating the swarming city: a MAS approach</p></title>
  <aug>
    <au><snm>Banos</snm><fnm>A</fnm></au>
    <au><snm>Chardonnel</snm><fnm>S</fnm></au>
    <au><snm>Lang</snm><fnm>C</fnm></au>
    <au><snm>Marilleau</snm><fnm>N</fnm></au>
    <au><snm>Th{\'e}venin</snm><fnm>T</fnm></au>
  </aug>
  <source>Proceedings of the 9th International Conference on Computers in Urban
  Planning and Urban Management</source>
  <pubdate>2005</pubdate>
  <fpage>29</fpage>
  <lpage>-30</lpage>
</bibl>

<bibl id="B24">
  <title><p>Geographically weighted regression</p></title>
  <aug>
    <au><snm>Brunsdon</snm><fnm>C</fnm></au>
    <au><snm>Fotheringham</snm><fnm>S</fnm></au>
    <au><snm>Charlton</snm><fnm>M</fnm></au>
  </aug>
  <source>Journal of the Royal Statistical Society: Series D (The
  Statistician)</source>
  <publisher>Wiley Online Library</publisher>
  <pubdate>1998</pubdate>
  <volume>47</volume>
  <issue>3</issue>
  <fpage>431</fpage>
  <lpage>-443</lpage>
</bibl>

<bibl id="B25">
  <title><p>SIMPOP: a multiagent system for the study of urbanism</p></title>
  <aug>
    <au><snm>Sanders</snm><fnm>L</fnm></au>
    <au><snm>Pumain</snm><fnm>D</fnm></au>
    <au><snm>Mathian</snm><fnm>H</fnm></au>
    <au><snm>Gu{\'e}rin Pace</snm><fnm>F</fnm></au>
    <au><snm>Bura</snm><fnm>S</fnm></au>
  </aug>
  <source>Environment and Planning B</source>
  <publisher>Pion Ltd</publisher>
  <pubdate>1997</pubdate>
  <volume>24</volume>
  <fpage>287</fpage>
  <lpage>-306</lpage>
</bibl>

<bibl id="B26">
  <title><p>Mod{\'e}lisation de la dynamique des syst{\`e}mes de peuplement: de
  SimpopLocal {\`a} SimpopNet.</p></title>
  <aug>
    <au><snm>Schmitt</snm><fnm>C</fnm></au>
  </aug>
  <source>PhD thesis</source>
  <publisher>Paris 1</publisher>
  <pubdate>2014</pubdate>
</bibl>

<bibl id="B27">
  <title><p>Space Matters: extending sensitivity analysis to initial spatial
  conditions in geosimulation models</p></title>
  <aug>
    <au><snm>Raimbault</snm><fnm>J</fnm></au>
    <au><snm>Cottineau</snm><fnm>C</fnm></au>
    <au><snm>Le Texier</snm><fnm>M</fnm></au>
    <au><snm>Le N{\'e}chet</snm><fnm>FL</fnm></au>
    <au><snm>Reuillon</snm><fnm>R</fnm></au>
  </aug>
  <source>Journal of Artificial Societies and Social Simulation</source>
  <publisher>Jasss</publisher>
  <pubdate>2019</pubdate>
  <volume>22</volume>
  <issue>4</issue>
</bibl>

<bibl id="B28">
  <title><p>Modeling social networks in geographic space: approach and
  empirical application</p></title>
  <aug>
    <au><snm>Arentze</snm><fnm>T</fnm></au>
    <au><snm>Berg</snm><fnm>P</fnm></au>
    <au><snm>Timmermans</snm><fnm>H</fnm></au>
  </aug>
  <source>Environment and Planning A</source>
  <publisher>SAGE Publications Sage UK: London, England</publisher>
  <pubdate>2012</pubdate>
  <volume>44</volume>
  <issue>5</issue>
  <fpage>1101</fpage>
  <lpage>-1120</lpage>
</bibl>

<bibl id="B29">
  <title><p>Interurban linkages through polynomially constrained distributed
  lags</p></title>
  <aug>
    <au><snm>Pigozzi</snm><fnm>BW</fnm></au>
  </aug>
  <source>Geographical Analysis</source>
  <pubdate>1980</pubdate>
  <volume>12</volume>
  <issue>4</issue>
  <fpage>340</fpage>
  <lpage>-352</lpage>
</bibl>

<bibl id="B30">
  <title><p>Urban gravity model based on cross-correlation function and Fourier
  analyses of spatio-temporal process</p></title>
  <aug>
    <au><snm>Chen</snm><fnm>Y</fnm></au>
  </aug>
  <source>Chaos, Solitons \& Fractals</source>
  <publisher>Elsevier</publisher>
  <pubdate>2009</pubdate>
  <volume>41</volume>
  <issue>2</issue>
  <fpage>603</fpage>
  <lpage>-614</lpage>
</bibl>

<bibl id="B31">
  <title><p>R{\'e}seaux et territoires-significations crois{\'e}es</p></title>
  <aug>
    <au><snm>Offner</snm><fnm>JM</fnm></au>
    <au><snm>Pumain</snm><fnm>D</fnm></au>
  </aug>
  <publisher>Editions de l'Aube</publisher>
  <pubdate>1996</pubdate>
</bibl>

<bibl id="B32">
  <title><p>Les "effets structurants" du transport: mythe politique,
  mystification scientifique</p></title>
  <aug>
    <au><snm>Offner</snm><fnm>JM</fnm></au>
  </aug>
  <source>Espace g{\'e}ographique</source>
  <pubdate>1993</pubdate>
  <volume>22</volume>
  <issue>3</issue>
  <fpage>233</fpage>
  <lpage>-242</lpage>
</bibl>

<bibl id="B33">
  <title><p>{Villes et r{\'e}seaux de transport : des interactions dans la
  longue dur{\'e}e, France, Europe, {\'E}tats-Unis}</p></title>
  <aug>
    <au><snm>Bretagnolle</snm><fnm>A</fnm></au>
  </aug>
  <source>PhD thesis</source>
  <publisher>Universit{\'e} Panth{\'e}on-Sorbonne - Paris I</publisher>
  <pubdate>2009</pubdate>
</bibl>

<bibl id="B34">
  <title><p>Caract{\'e}risation et mod{\'e}lisation de la co-{\'e}volution des
  r{\'e}seaux de transport et des territoires</p></title>
  <aug>
    <au><snm>Raimbault</snm><fnm>J</fnm></au>
  </aug>
  <source>PhD thesis</source>
  <publisher>Universit{\'e} Paris 7 Denis Diderot</publisher>
  <pubdate>2018</pubdate>
</bibl>

<bibl id="B35">
  <title><p>Hierarchy in cities and city systems</p></title>
  <aug>
    <au><snm>Batty</snm><fnm>M</fnm></au>
  </aug>
  <source>Hierarchy in natural and social sciences</source>
  <publisher>Dordrecht: Springer</publisher>
  <pubdate>2006</pubdate>
  <fpage>143</fpage>
  <lpage>-168</lpage>
</bibl>

<bibl id="B36">
  <title><p>Calibration of a density-based model of urban
  morphogenesis</p></title>
  <aug>
    <au><snm>Raimbault</snm><fnm>J</fnm></au>
  </aug>
  <source>PloS one</source>
  <publisher>Public Library of Science</publisher>
  <pubdate>2018</pubdate>
  <volume>13</volume>
  <issue>9</issue>
  <fpage>e0203516</fpage>
</bibl>

<bibl id="B37">
  <title><p>Eurostat Geographical Data</p></title>
  <aug>
    <au><cnm>EUROSTAT</cnm></au>
  </aug>
  <source>\url{http://ec.europa.eu/eurostat/web/gisco}</source>
  <pubdate>2014</pubdate>
</bibl>

<bibl id="B38">
  <title><p>Multi-modeling the morphogenesis of transportation
  networks</p></title>
  <aug>
    <au><snm>Raimbault</snm><fnm>J</fnm></au>
  </aug>
  <source>Artificial Life Conference Proceedings</source>
  <pubdate>2018</pubdate>
  <fpage>382</fpage>
  <lpage>-383</lpage>
</bibl>

<bibl id="B39">
  <title><p>Rules for Biologically Inspired Adaptive Network Design</p></title>
  <aug>
    <au><snm>Tero</snm><fnm>A</fnm></au>
    <au><snm>Takagi</snm><fnm>S</fnm></au>
    <au><snm>Saigusa</snm><fnm>T</fnm></au>
    <au><snm>Ito</snm><fnm>K</fnm></au>
    <au><snm>Bebber</snm><fnm>DP</fnm></au>
    <au><snm>Fricker</snm><fnm>MD</fnm></au>
    <au><snm>Yumiki</snm><fnm>K</fnm></au>
    <au><snm>Kobayashi</snm><fnm>R</fnm></au>
    <au><snm>Nakagaki</snm><fnm>T</fnm></au>
  </aug>
  <source>Science</source>
  <pubdate>2010</pubdate>
  <volume>327</volume>
  <issue>5964</issue>
  <fpage>439</fpage>
  <lpage>442</lpage>
</bibl>

<bibl id="B40">
  <title><p>Mathematics and morphogenesis of cities: A geometrical
  approach</p></title>
  <aug>
    <au><snm>Courtat</snm><fnm>T</fnm></au>
    <au><snm>Gloaguen</snm><fnm>C</fnm></au>
    <au><snm>Douady</snm><fnm>S</fnm></au>
  </aug>
  <source>Physical Review E</source>
  <pubdate>2011</pubdate>
  <volume>83</volume>
  <issue>3</issue>
  <fpage>036106</fpage>
</bibl>

<bibl id="B41">
  <title><p>Multi-dimensional Urban Network Percolation</p></title>
  <aug>
    <au><snm>Raimbault</snm><fnm>J</fnm></au>
  </aug>
  <source>Journal of Interdisciplinary Methods and Issues in Science. In
  press</source>
  <pubdate>2019</pubdate>
</bibl>

<bibl id="B42">
  <title><p>De la forme urbaine {\`a} la structure m{\'e}tropolitaine: une
  typologie de la configuration interne des densit{\'e}s pour les principales
  m{\'e}tropoles europ{\'e}ennes de l'Audit Urbain</p></title>
  <aug>
    <au><snm>Le N{\'e}chet</snm><fnm>F</fnm></au>
  </aug>
  <source>Cybergeo: European Journal of Geography</source>
  <pubdate>2015</pubdate>
</bibl>

<bibl id="B43">
  <title><p>Towards new metrics for urban road networks: Some preliminary
  evidence from agent-based simulations</p></title>
  <aug>
    <au><snm>Banos</snm><fnm>A</fnm></au>
    <au><snm>Genre Grandpierre</snm><fnm>C</fnm></au>
  </aug>
  <source>Agent-based models of geographical systems</source>
  <publisher>Dordrecht: Springer</publisher>
  <pubdate>2012</pubdate>
  <fpage>627</fpage>
  <lpage>-641</lpage>
</bibl>

<bibl id="B44">
  <title><p>OpenMOLE, a workflow engine specifically tailored for the
  distributed exploration of simulation models</p></title>
  <aug>
    <au><snm>Reuillon</snm><fnm>R</fnm></au>
    <au><snm>Leclaire</snm><fnm>M</fnm></au>
    <au><snm>Rey Coyrehourcq</snm><fnm>S</fnm></au>
  </aug>
  <source>Future Generation Computer Systems</source>
  <pubdate>2013</pubdate>
  <volume>29</volume>
  <issue>8</issue>
  <fpage>1981</fpage>
  <lpage>-1990</lpage>
</bibl>

<bibl id="B45">
  <title><p>NetLogo</p></title>
  <aug>
    <au><snm>Wilensky</snm><fnm>U</fnm></au>
  </aug>
  <pubdate>1999</pubdate>
</bibl>

<bibl id="B46">
  <title><p>An urban morphogenesis model capturing interactions between
  networks and territories</p></title>
  <aug>
    <au><snm>Raimbault</snm><fnm>J</fnm></au>
  </aug>
  <source>The Mathematics of Urban Morphology</source>
  <publisher>Cham: Springer</publisher>
  <pubdate>2019</pubdate>
  <fpage>383</fpage>
  <lpage>-409</lpage>
</bibl>

<bibl id="B47">
  <title><p>An introduction to econophysics: correlations and complexity in
  finance</p></title>
  <aug>
    <au><snm>Mantegna</snm><fnm>RN</fnm></au>
    <au><snm>Stanley</snm><fnm>HE</fnm></au>
  </aug>
  <publisher>Cambridge: Cambridge university press</publisher>
  <pubdate>2000</pubdate>
</bibl>

<bibl id="B48">
  <title><p>{Financial Applications of Random Matrix Theory: a short
  review}</p></title>
  <aug>
    <au><snm>{Bouchaud}</snm><fnm>J. P.</fnm></au>
    <au><snm>{Potters}</snm><fnm>M.</fnm></au>
  </aug>
  <source>ArXiv e-prints</source>
  <pubdate>2009</pubdate>
</bibl>

<bibl id="B49">
  <title><p>{Levels of complexity in financial markets}</p></title>
  <aug>
    <au><snm>{Bonanno}</snm><fnm>G.</fnm></au>
    <au><snm>{Lillo}</snm><fnm>F.</fnm></au>
    <au><snm>{Mantegna}</snm><fnm>R. N.</fnm></au>
  </aug>
  <source>Physica A Statistical Mechanics and its Applications</source>
  <pubdate>2001</pubdate>
  <volume>299</volume>
  <fpage>16</fpage>
  <lpage>27</lpage>
</bibl>

<bibl id="B50">
  <title><p>A tool for filtering information in complex systems</p></title>
  <aug>
    <au><snm>Tumminello</snm><fnm>M</fnm></au>
    <au><snm>Aste</snm><fnm>T</fnm></au>
    <au><snm>Di Matteo</snm><fnm>T</fnm></au>
    <au><snm>Mantegna</snm><fnm>RN</fnm></au>
  </aug>
  <source>Proceedings of the National Academy of Sciences of the United States
  of America</source>
  <publisher>National Acad Sciences</publisher>
  <pubdate>2005</pubdate>
  <volume>102</volume>
  <fpage>10421</fpage>
  <lpage>-10426</lpage>
</bibl>

<bibl id="B51">
  <title><p>Multivariate realised kernels: consistent positive semi-definite
  estimators of the covariation of equity prices with noise and non-synchronous
  trading</p></title>
  <aug>
    <au><snm>Barndorff Nielsen</snm><fnm>OE</fnm></au>
    <au><snm>Hansen</snm><fnm>PR</fnm></au>
    <au><snm>Lunde</snm><fnm>A</fnm></au>
    <au><snm>Shephard</snm><fnm>N</fnm></au>
  </aug>
  <source>Journal of Econometrics</source>
  <publisher>Elsevier</publisher>
  <pubdate>2011</pubdate>
  <volume>162</volume>
  <fpage>149</fpage>
  <lpage>-169</lpage>
</bibl>

<bibl id="B52">
  <title><p>Wavelets in economics and finance: Past and future</p></title>
  <aug>
    <au><snm>Ramsey</snm><fnm>JB</fnm></au>
  </aug>
  <source>Studies in Nonlinear Dynamics \& Econometrics</source>
  <pubdate>2002</pubdate>
  <volume>6</volume>
</bibl>

<bibl id="B53">
  <title><p>Apparent multifractality in financial time series</p></title>
  <aug>
    <au><snm>Bouchaud</snm><fnm>J P</fnm></au>
    <au><snm>Potters</snm><fnm>M</fnm></au>
    <au><snm>Meyer</snm><fnm>M</fnm></au>
  </aug>
  <source>The European Physical Journal B-Condensed Matter and Complex
  Systems</source>
  <publisher>Springer</publisher>
  <pubdate>2000</pubdate>
  <volume>13</volume>
  <issue>3</issue>
  <fpage>595</fpage>
  <lpage>-599</lpage>
</bibl>

<bibl id="B54">
  <title><p>In Honor of the Nobel Laureates Robert C. Merton and Myron S.
  Scholes: A Partial Differential Equation that Changed the World</p></title>
  <aug>
    <au><snm>Jarrow</snm><fnm>RA</fnm></au>
  </aug>
  <source>The Journal of Economic Perspectives</source>
  <publisher>JSTOR</publisher>
  <pubdate>1999</pubdate>
  <fpage>229</fpage>
  <lpage>-248</lpage>
</bibl>

<bibl id="B55">
  <title><p>MTS: All-Purpose Toolkit for Analyzing Multivariate Time Series
  (MTS) and Estimating Multivariate Volatility Models</p></title>
  <aug>
    <au><snm>Tsay</snm><fnm>RS</fnm></au>
  </aug>
  <pubdate>2015</pubdate>
  <url>http://CRAN.R-project.org/package=MTS</url>
  <note>R package version 0.33</note>
</bibl>

<bibl id="B56">
  <title><p>Quality assessment of the French OpenStreetMap dataset</p></title>
  <aug>
    <au><snm>Girres</snm><fnm>JF</fnm></au>
    <au><snm>Touya</snm><fnm>G</fnm></au>
  </aug>
  <source>Transactions in GIS</source>
  <publisher>Wiley Online Library</publisher>
  <pubdate>2010</pubdate>
  <volume>14</volume>
  <issue>4</issue>
  <fpage>435</fpage>
  <lpage>-459</lpage>
</bibl>

<bibl id="B57">
  <title><p>A nested factor model for non-linear dependencies in stock
  returns</p></title>
  <aug>
    <au><snm>Chicheportiche</snm><fnm>R</fnm></au>
    <au><snm>Bouchaud</snm><fnm>J P</fnm></au>
  </aug>
  <source>Quantitative Finance</source>
  <publisher>Taylor \& Francis</publisher>
  <pubdate>2015</pubdate>
  <volume>15</volume>
  <issue>11</issue>
  <fpage>1789</fpage>
  <lpage>-1804</lpage>
</bibl>

<bibl id="B58">
  <title><p>Beyond Corroboration: Strengthening Model Validation by Looking for
  Unexpected Patterns</p></title>
  <aug>
    <au><snm>Ch{\'e}rel</snm><fnm>G</fnm></au>
    <au><snm>Cottineau</snm><fnm>C</fnm></au>
    <au><snm>Reuillon</snm><fnm>R</fnm></au>
  </aug>
  <source>PLoS ONE</source>
  <pubdate>2015</pubdate>
  <volume>10</volume>
  <issue>9</issue>
  <fpage>e0138212</fpage>
</bibl>

<bibl id="B59">
  <title><p>{French Roadmap for complex Systems 2008-2009}</p></title>
  <aug>
    <au><snm>{Bourgine}</snm><fnm>P.</fnm></au>
    <au><snm>{Chavalarias}</snm><fnm>D.</fnm></au>
    <au><cnm>al.</cnm></au>
  </aug>
  <source>arXiv preprint arXiv:0907.2221</source>
  <pubdate>2009</pubdate>
</bibl>

</refgrp>
} 

\end{document}